\documentstyle[twocolumn,aps]{revtex}

\input{epsf}

\begin{document}
\author{D. Blume$^{(1)}$, B. D. Esry$^{(2)}$, Chris H. Greene$^{(3)}$, 
N. N. Klausen$^{(3)}$, and G. J. Hanna$^{(1)}$} 
\title{Formation of atomic tritium clusters and condensates}
\date{\today }
\address{$^{(1)}$Department of Physics, Washington State University, 
Pullman, Washington 99164-2814, USA;\\
$^{(2)}$Department of Physics, Kansas State University,
Manhattan, Kansas  66506, USA;\\
$^{(3)}$Department of Physics and JILA, University of Colorado, 
Boulder, Colorado 80309-0440, USA\\
}
\maketitle

\begin{abstract}
We present an extensive study of the static and dynamic properties of 
systems of spin-polarized tritium atoms.  In particular, we calculate the 
two-body 
$|F,m_F \rangle = |0,0 \rangle$
$s$-wave scattering length and show that it can be manipulated
via a Feshbach resonance at a field strength of about 870~G.  Such a 
resonance might be exploited to make and 
control a Bose-Einstein condensate of tritium
in the $|0,0 \rangle$ state.
It is further shown that the quartet tritium trimer is the only bound hydrogen
isotope and that its single vibrational bound state is a Borromean state.
The ground state properties of larger spin-polarized tritium clusters are also
presented and compared with those of helium clusters.
\end{abstract}

\draft
\pacs{05.30.Jp,34.50,36.40.-c}


In 1976~\cite{stwa76}, Stwalley and Nosanow 
suggested, based on statistical arguments,  
that the system of spin-polarized bosonic tritium 
atoms
behaves ``very much
like $^4$He''.  
To the best of our knowledge, their arguments have not yet been
tested by a microscopic quantum mechanical treatment. 
A detailed theoretical study of spin-polarized tritium systems,
namely spin-polarized atomic tritium clusters and optically-pumped 
tritium condensates, is the objective of the present work.  In this Letter,
we present results for both structural and scattering properties of tritium dimers,
trimers, and clusters.

Pioneering experimental studies of the lowest quartet state
of spin-polarized atomic trimers have been pursued recently
for sodium and potassium~\cite{higg96reho01a}.  Study of these trimers, which 
were prepared on the surface of large $^4$He clusters,
revealed that three-body effects are surprisingly important~\cite{higg00}. 
We are not aware, though, of any experimental or theoretical studies of 
larger spin-polarized atomic cluster systems. 
Bosonic helium systems --- i.e. liquid bulk $^4$He, two-dimensional
$^4$He films, and finite size $^4$He$_N$ 
clusters~\cite{whal94} ---
have, of course, been studied extensively.  This Letter thus
presents the first predictions for spin-polarized atomic clusters.
In particular, we characterize spin-polarized tritium 
clusters [in the following denoted by (T$\uparrow$)$_N$] with up
to $N=40$ tritium atoms, and compare their energetic
and structural properties with those of bosonic $^4$He$_N$ clusters.
We hope that this study will stimulate further experimental work.
Of particular interest is the lowest quartet state of the tritium trimer,
which we predict to be a Borromean or halo state, and tritium
cluster formation in the presence of an external magnetic field.

We also point out the possibility for creating an optically-pumped
gaseous tritium condensate. 
Bose-Einstein condensates (BECs) are, to first order,
well characterized by the two-body
$s$-wave scattering length between two atoms.
It will be shown below that the triplet two-body
$s$-wave scattering length $a_t$ of two tritium
atoms is large and negative, implying an unstable
condensate of spin-polarized tritium atoms.
We find, though, that there is an unusually
broad Feshbach resonance~\cite{cour98inou98} 
for two high-field-seeking, $|F,m_F \rangle = |0,0 \rangle$ tritium atoms 
($F$ denotes the total angular momentum, and $m_F$ the
magnetic quantum number of this state).

While condensation of atomic hydrogen was 
realized experimentally in 1998~\cite{frie98}, it was a difficult
experiment --- at least in part because of hydrogen's small triplet
scattering length that limits the utility
of evaporative cooling.  Nevertheless, owing to
hydrogen's simplicity, it remains an important species to study.
For instance, properties such as the interatomic potential and spin 
relaxation rates can be obtained theoretically from first principles. 
Unfortunately, we find no Feshbach resonance at reasonable field strengths
for hydrogen.  Thus, the resonance for tritium may permit faster 
condensation 
of a hydrogen-like atom, and allow for the formation of a stable
BEC of tritium atoms with controllable properties
in an optical dipole trap~\cite{barr01}. 
Formation of such a tritium condensate should enhance the lively
interplay between theory and experiment.

The behaviour of atomic tritium clusters and condensates is
primarily determined by the two-body interaction potential for two 
tritium atoms, which is identical to that for two H or D 
atoms, except for the isotope-dependent adiabatic correction.
Since there are only two electrons, these 
dimers are among the
few for which highly accurate {\em{ab initio}} potentials 
are available. In the following, we
concentrate on the singlet ground state ($S=0$, $X ^1 \Sigma _g ^+$), 
and on the triplet ground state ($S=1$, $b ^3 \Sigma _u ^+$) of the
tritium dimer. 

To construct the two-body $S=0$ and $S=1$ Born-Oppenheimer 
interaction potentials for hydrogen and its heavier isotopes,
highly accurate {\em{ab initio}} data for the short
range part~\cite{kolo74kolo90jami00} that
incorporate the mass-dependent adiabatic correction 
are connected smoothly with an analytical expression
describing the long-range behavior~\cite{bukt74,herr64,yan96}. 
This procedure results in six potential curves describing 
H$_2$, D$_2$, and T$_2$ in their $S=0$ and $S=1$ states, respectively,
which are then used in the radial Schr\"odinger equation
describing the relative motion of a particle
with reduced mass $m/2$.  The mass dependence of 
the singlet and triplet scattering lengths has been discussed extensively
in the literature\cite{will93,jami98jami01}. 
Here, we estimate the uncertainty of our two-body scattering observables
by solving the radial Schr\"odinger equation using both
the reduced atomic mass and the reduced nuclear mass for each two-body potential
described above.

\vspace*{-1in}
\begin{figure}
\centerline{\epsfxsize=2.5in\epsfbox{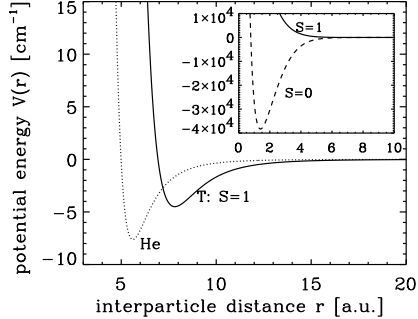}}
\vspace*{-0.05in}
\caption{Tritium dimer triplet  $b^3 \Sigma _u ^+$
potential (solid line) together with the He dimer 
potential (dotted line) as a function of the interparticle distance $r$.
The inset compares the tritium dimer triplet potential ($S=1$, solid line)
with the tritium  dimer singlet ground state potential ($S=0$, dashed line). 
Note the different vertical scales of the main figure and the inset.}
\label{fig_potential}
\end{figure}
Figure~\ref{fig_potential} 
compares the (T$\uparrow$)$_2$ potential 
(solid line) with the similarly shallow He$_2$ ground state potential,
LM2M2 from Aziz and Slaman~\cite{aziz91} (dotted line).
The T$\uparrow$ dimer and the He dimer potentials have a well depth of
$D_e=-4.6$cm$^{-1}$ and $-7.6$cm$^{-1}$, respectively.
Note that on the scale shown in Fig.~\ref{fig_potential}, 
the H$\uparrow$ and D$\uparrow$ dimer potentials would be
indistinguishable from the T$\uparrow$ dimer potential.
The minimum of the (T$\uparrow$)$_2$ potential
lies at a significantly larger interparticle distance
($r_e=7.8$a.u.) than for the He dimer ($r_e=5.6$a.u.). 
Given the lighter mass of the tritium atom ($m=5496.9$a.u.)
compared to that of the $^4$He atom ($m=7296.3$a.u.),
it is not surprising that the T$\uparrow$ dimer is not bound,
even though the tritium van der Waals coefficient $C_6=6.499$a.u.
is larger than that for He, $C_6=1.367$a.u.
(recall that the $^4$He dimer binding energy is 
only $-9.1 \times 10^{-4}$cm$^{-1}$~\cite{janz95}). 
For comparison,
the inset of Fig.~\ref{fig_potential} shows the tritium triplet 
potential (solid line) together 
with the tritium singlet 
potential (dashed line).
The singlet curve is almost four orders of magnitude
deeper than the triplet curve and supports 27
vibrational $s$-wave bound states.

In agreement with values tabulated in the literature~\cite{will93},
we calculate the
two-body $s$-wave triplet scattering lengths $a_t$ for (H$\uparrow$)$_2$
to be 
$a_t=1.33$a.u. [$1.33$a.u.], and for (D$\uparrow$)$_2$ to
be $-6.89$a.u. [$-6.88$a.u.],
using the reduced atomic [reduced nuclear]
mass.  From symmetry considerations, the $s$-wave
scattering length for (D$\uparrow$)$_2$ is not an observable,
and is given here for diagnostic purposes only.
For tritium, we predict a positive singlet scattering length,
$a_s = 34.6$a.u. [$35.8$a.u.], 
and a large negative 
$s$-wave scattering length, $a_t=-82.1$a.u. [$-81.9$a.u.].
The T$\uparrow$ dimer does not possess a bound state,
but this large negative $a_t$ indicates that it
is only ``slightly short of binding''.

Despite the fact that the tritium triplet scattering length is
negative --- implying an unstable condensate --- it may be possible
to form a stable tritium condensate utilizing a Feshbach resonance.
Coupled-channel scattering calculations that couple the
singlet and triplet subspaces reveal such a Feshbach resonance,
i.e. a diverging scattering length for two atoms characterized by 
quantum numbers $F$ and $m_F$, as a function of the magnetic field strength.
The coupling arises through the atomic 
hyperfine interaction, and has to be accounted for by an effective
two-atom Hamiltonian~\cite{burk98}.
Feshbach resonances have been observed experimentally
for $^{23}$Na and $^{85}$Rb~\cite{cour98inou98} among others. 
The latter paved the way for a series of important BEC experiments, some of
which enter the large interaction regime~\cite{corn00} while others
probe the collapse 
regime~\cite{donl01}.

\vspace*{-0.95in}
\begin{figure}
\centerline{\epsfxsize=2.5in\epsfbox{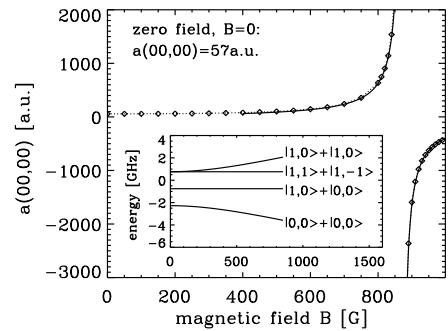}}
\vspace*{-0.05in}
\caption{$a(00+00)$ scattering length (diamonds;
using the reduced atomic mass in 
the coupled-channel calculation,
see text) as 
a function of the magnetic field strength $B$ 
(dotted lines are shown to guide the eye). The solid line 
describes the behavior for the range
$B\in [400,1300\mbox{G}]$ well using the following parameterization,
$a(00,00) = a_{BG}(1-\Delta/(B-B_R))$ with $a_{BG}=-37.9$a.u., 
$\Delta=-1238$G,
and $B_R=870.8$G; however, 
the fit is inaccurate at low fields.
Inset: Threshold energies in GHz as a function of magnetic 
field $B$ in Gauss. 
The assignment of quantum numbers is approximate, except for
$| 1,1 \rangle + | 1,-1 \rangle$, which is an exact eigenstate without
admixtures.}
\label{fig_feshbach}
\end{figure}
We find that collisions between two tritium atoms,
each in their $|F, m_F \rangle = |0,0 \rangle$ state, 
results in a scattering lenth of 
$a(00,00)=57$a.u. [64a.u.] for zero magnetic field
strength $B$, using the atomic [nuclear] mass.
As $B$ increases, the scattering length $a(00+00)$ rises 
and eventually goes through infinity
across a broad Feshbach resonance centered at $B=870$G [810G]
(see Fig.~\ref{fig_feshbach}).  We also looked for, but could not find, a
similar resonance for collisions of hydrogen atoms.  Note
that our predictions are not 
sensitive to a replacement of the atomic mass by the nuclear mass.

Formation of a tritium condensate in the high-field-seeking $|0,0\rangle$ state would require
some variety of nonmagnetic trap, such as the dipole CO$_2$ laser trap
that has already produced a $^{87}$Rb condensate~\cite{barr01}.  Since
the electric dipole polarizability of tritium is only 4.5~a.u., the 12~W
CO$_2$ laser setup of~\cite{barr01} would only produce trap depths of
the order of a few microkelvin.  There appears to be no reason why much
stronger CO$_2$ lasers could not be utilized, however.
Realistically,
the formation of a tritium condensate will probably require laser
intensities at least an order of magnitude more intense.  The large
magnitude of the zero-field scattering length would make evaporative
cooling far more effective than is the case for spin-polarized
hydrogen.  Another possible way to cool the spinless substate of tritium
would be to implement a recent proposal to cool an atomic gas through
magnetic field ramps across a Feshbach resonance~\cite{dunn02}.  Despite
these technical difficulties to be overcome, recent improvements in
trapping and cooling technology would appear to make the creation of a
tritium condensate a viable possibility.

To determine the bound state properties of (T$\uparrow$)$_N$
clusters with $N>2$, we first have to investigate the importance of
non-additive contributions to the many-body interaction potential.
Although three-body contributions,
i.e., the Axilrod-Teller term~\cite{axil43muto43} and
three-body exchange terms~\cite{tang98}, 
are significant for the hydrogen
trimer in its electronic ground state, they should be less important
for the spin-polarized trimer since the
classical atom-atom equilibrium distance of the quartet trimer
is more than five times 
as large as that for the doublet ground state trimer.
Our calculations show that inclusion of 
the
damped
Axilrod-Teller term~\cite{sach00}
raises the ground state energy of the spin-polarized tritium 
trimer by roughly 1.6\%, and that of the larger clusters slightly more,
e.g., by about 6~\% for $N=40$.
To describe (T$\uparrow$)$_N$ clusters,
we 
assume in the following a simple pairwise additive potential energy
surface, $V=\sum_{i<j} ^N V(r_{ij})$,
where $V(r)$ denotes the triplet $b^3 \Sigma ^+ _u$ two-body potential.
Conceivably, a more sophisticated many-body potential energy surface,
which includes effects beyond the two-body potential, could modify
our quantitative results somewhat,
but we do not expect qualitative changes.

For the T$\uparrow$ trimer, we use the adiabatic hyperspherical representation
~\cite{esry99bblum00a}.  Including only one adiabatic
channel yields a single bound state  with energy 
$-1.60\times 10^{-3}$~cm$^{-1}$.
Coupling 25 adiabatic channels results in an energy of $-3.19\times 10^{-3}$~cm$^{-1}$
with an uncertainty of 10$^{-5}$~cm$^{-1}$ and still no excited states.  In the limit
that an infinite number of channels are coupled, the bound state energy becomes 
exact, so the uncertainty is the result of including a finite number of channels.
Upon inclusion of the Axilrod-Teller three-body term, the ground state energy is raised
to $-3.14\times 10^{-3}$~cm$^{-1}$ --- again with uncertainty in the last digit.
There is thus a single $L=0$ bound state for quartet tritium ($L$ is the total orbital angular momentum);
no $L>0$ bound states are expected since none exist for the $^4$He trimer~\cite{lee01}.

Since the T$\uparrow$ dimer is unbound, the T$\uparrow$ trimer
is a Borromean state~\cite{fedo94fedo94afedo95}. 
One may then ask: does the T$\uparrow$ trimer state have Efimov character~\cite{efim70}?
To investigate this question, we apply a simple quantitative criterion~\cite{esry96}, although others exist~\cite{braa02}.
If the bound state disappears when one makes the potential more attractive 
(here achieved by simply multiplying the two-body potential with 
an overall scaling factor greater than 1), 
then the state under investigation is an Efimov state; if the bound state
does not disappear, it is not an Efimov state.
Applying this criterion, our coupled-channel calculations indicate
that the bound T$\uparrow$ trimer state is not an Efimov state.
In short, the T$\uparrow$ trimer has exactly one $L=0$  bound state,
a Borromean state that is highly diffuse spatially. 

To calculate the energetics and structural properties of 
(T$\uparrow$)$_N$ clusters with up to $N=40$ atoms, we employ
the diffusion quantum Monte Carlo (DMC) technique~\cite{hamm94}. This method
solves the time-independent many-body Schr\"odinger equation essentially
exactly, to within a statistical error.
Here, we employ the DMC method with importance sampling~\cite{hamm94},
using a descendant weighting scheme~\cite{kalo67liu74barn91} 
for the extrapolation of structural properties.  Our
guiding wave functions~\cite{hamm94}, 
which enter the DMC calculation, have the analytical form given in 
Eq.~(5) of Ref.~\cite{lewe97}, and recover between 84 and
96\% of the DMC energy when used in a variational 
quantum Monte Carlo calculation.
The DMC technique as implemented here determines
only the ground state of the system;
accessing excited state properties is in general not
possible without algorithmic modifications.

\vspace*{-2.15in}
\begin{figure}
\hspace*{0.1in}
\centerline{\epsfxsize=5.in\epsfbox{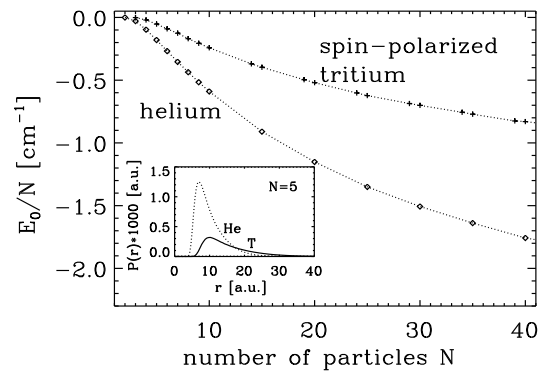}}
\vspace*{-1.7in}
\caption{Ground state energy per
particle $E_0/N$ for (T$\uparrow$)$_N$ clusters (pluses) and $^4$He$_N$ 
clusters (diamonds)
as a function of $N$. Dotted lines are shown to guide the eye.
Inset: Pair distribution $P(r)$ for 
(T$\uparrow$)$_5$ cluster (solid line) and $^4$He$_5$ 
cluster (dotted line), calculated by the DMC technique
with importance sampling. $P(r)$ is normalized
such that $\int _ 0 ^{\infty} P(r) r^2 \, dr =1$.
}
\label{fig_energy}
\end{figure}
For comparison, we find a 
DMC binding energy of $-2.9(5) \times 10^{-3}$cm$^{-1}$ for the 
T$\uparrow$ trimer,
in agreement with our hyperspherical calculation.
The number in brackets denotes the statistical uncertainty.
Figure~\ref{fig_energy} 
compares the ground state energy per particle $E_0/N$
of (T$\uparrow$)$_N$ clusters (pluses) 
with those of $^4$He$_N$ clusters (using the LM2M2 potential \cite{aziz91}, diamonds).
This shows that
(T$\uparrow$)$_N$ clusters are even more 
weakly bound than $^4$He$_N$ clusters with the same number of atoms.

The comparison between (T$\uparrow$)$_N$  and
$^4$He$_N$ clusters can be extended by considering their structural properties.
For example, we find that the (T$\uparrow$)$_5$ cluster
has an average interparticle distance of $ \langle r_{ij} \rangle$ of $22.2$a.u.;
the $^4$He$_5$ cluster, on the other hand, is $ \langle r_{ij} \rangle=13.6$a.u.
Even if one takes into account that the classical equilibrium distance of the
tritium triplet potential is about 2.2a.u. larger than that of the
He dimer potential, the difference between the 
expectation values of the interparticle distance for these $N=5$ clusters
indicates that the spin-polarized tritium system is even more 
diffuse than the $^4$He$_5$ cluster.
To illustrate this aspect further, the inset of Fig.~\ref{fig_energy}
compares the pair distribution of the (T$\uparrow$)$_5$ cluster
(solid line) with that of the $^4$He$_5$ cluster (dotted line).
Clearly, the pair distribution of the (T$\uparrow$)$_5$ cluster
is much broader than that of the $^4$He$_5$ cluster.
We find similar behavior for clusters with more particles.

The tritium trimer is the smallest spin-polarized cluster (the dimer is unbound). 
As discussed above, the two-body potential 
for spin-polarized hydrogen is almost identical to that of tritium.
The hydrogen atom, however, is about a factor of three lighter.
Consequently, there is no bound state for the H$\uparrow$ trimer.
An interesting question to ask is the following: how many atoms are needed to form a bound system 
of spin-polarized hydrogen atoms? 
Initial exploratory studies show that more than 100 atoms are needed
to form a bound spin-polarized hydrogen cluster. 
The smallest spin-polarized hydrogen cluster could then be 
thought of as a ``super-Borromean''
cluster for which all smaller subsystems are unbound.
A detailed study will be published elsewhere.

In summary, this Letter proposes a number of intriguing possibilities
for the physics of tritium systems.  We point out the possibility for
forming a tritium condensate with controllable parameters via a Feshbach
resonance.  Down the road, one can imagine trapping an atomic hydrogen gas together
with an atomic tritium gas, or possibly including deuterium to study fermion
systems.  We further found that the spin-polarized trimer possesses a Borromean,
or halo, state.  In addition, we mapped out the properties of larger ``exotic''
spin-polarized tritium clusters.  Studies of spin-polarized clusters are 
interesting by themselves~\cite{hott02}, 
as they enter new many-body physics regimes.
For instance, unexpected physics may emerge from manipulating the two-body
interaction via the Fesh\-bach resonance in a cluster.  In short, tritium
offers a wealth of interesting physics by virtue of its weak 
attraction.


{\em Acknowledgments:} This work was supported by the National
Science Foundation. BDE acknowledges support from the Research
Corporation.  We thank V. Kokoouline for help
with the two-body coupled-channel calculations.
Fruitful discussions with L.~W. Bruch
at an early stage of this project are acknowledged.

\end{document}